\documentclass[prd, aps, reprint, amsfonts, amssymb, amsmath, showpacs, nofootinbib, superscriptaddress, preprintnumbers, onecolumn, notitlepage ]{revtex4-1}
\usepackage{graphicx, multirow,soul}
\usepackage{hyperref}
\usepackage[all]{hypcap}
\usepackage{url}
\usepackage{minibox}
\usepackage{color}
\usepackage{subfigure}
\usepackage{slashed}
\usepackage{float}
\usepackage{amssymb}

\newcommand{\order}[1]{\mathcal{O}\left(#1\right)}

\begin{document}

\title{On stochastically sampling color configurations}

\author{Joshua Isaacson}\email{isaacson@fnal.gov}
\affiliation{Theoretical Physics Department, Fermi National Accelerator Laboratory, P.O. Box 500, Batavia, IL 60510, USA.}
\author{Stefan Prestel}\email{sprestel@fnal.gov}
\affiliation{Theoretical Physics Department, Fermi National Accelerator Laboratory, P.O. Box 500, Batavia, IL 60510, USA.}

\begin{abstract}
Parton shower algorithms are key components of theoretical predictions for 
high-energy collider physics. Work towards more accurate parton shower 
algorithms is thus pursued along many different avenues.
The systematic treatment of subleading color corrections in parton shower
algorithms is however technically challenging and remains elusive. 
In this article, we present an efficient and numerically stable algorithm to sample 
color configurations at fixed $N_C=3$, using the correct color factor including
subleading corrections with a parton shower. The algorithm is implemented as
stand-alone program that can be interfaced to the \textsc{Pythia} event generator.
Preliminary comparisons to to LEP data are presented.
\end{abstract}

\preprint{\minibox[]{FERMILAB--PUB--18--296--T}}
\maketitle

\section{Introduction}\label{sec:introduction}

\noindent
High energy physics event generators are tools that enable comparisons
of theory calculations to measurements at high-energy collider experiments~\cite{Buckley:2011ms}.
This is achieved by simulating scattering events in as much detail
as possible. Event generators are thus heavily
used in analysis prototyping, to compare precision Standard-Model (SM) measurements
to theory, or to derive indirect bounds on new phenomena. This is only
possible if event generators incorporate precise calculations in perturbative
QCD\@. Dedicated efforts over the last decade have ensured that scattering
events with well-separated jets are described with high accuracy. The accuracy of
the modeling of jet structure and evolution is much less understood. Jet
evolution can be generated by parton showers~\cite{Fox:1979ag,*Sjostrand:1985xi,*Webber:1986mc,*Gustafson:1987rq} -- programs that are based on 
solving the leading order DGLAP evolution equations~\cite{Gribov:1972ri,*Dokshitzer:1977sg,*Altarelli:1977zs} by explicitly 
constructing states with multiple branchings. Several approximations
are then made to allow an iterative, probabilistic interpretation of these 
evolution equations. Precision SM measurements or indirect limit setting using 
jet substructure will
mean that some if not all of the parton shower approximations will have to be
reevaluated and removed.

This article describes an algorithm to overcome the $N_C \to \infty$ limit
(also known as the Leading Color (LC)
approximation) typically employed by parton shower programs.
In the past few years, there has been a resurgence in the desire to 
understand the formalism and requirements for full-color parton 
showers~\cite{Platzer:2012np,Bellm:2018wwz,Nagy:2015hwa,Nagy:2012bt,
Nagy:2007ty,Martinez:2018ffw}. 
The goal of the current work is to present a realistic algorithm to efficiently and
consistently implement a means to sample, according to single-parton insertion
operators, arbitrary color configurations defined
with $N_C=3$ within a parton shower. We introduce a new parton shower algorithm
that strictly employs $N_C=3$ and can be matched onto non-perturbative 
hadronization models, which does not experience factorial growth in the time 
to process each subsequent emission that was previously observed~\cite{Platzer:2012np}.
The manuscript first introduces some basic parton shower concepts in 
Sec.~\ref{sec:ps_basics} before describing the new $N_C=3$ parton shower algorithm in
Sec.~\ref{sec:Algorithm}. This is followed by results and comparisons to
experimental measurements in Sec.~\ref{sec:results}, before giving conclusions
and an outlook in Sec.~\ref{sec:conclusions}.
 
\section{Parton Shower Basics}\label{sec:ps_basics}

\noindent
Parton showers are a crucial component of event simulations, as they link
low-multiplicity hard scatterings calculations to realistic jet observables 
by explicitly simulating perturbative jet formation and evolution. This is
achieved by solving the leading-order DGLAP-like renormalization group equations
\begin{equation}\label{eq:pdf_evolution}
  \frac{{\rm d}\,f_{a}(x,t)}{{\rm d}\ln t}=
  \sum_{b=q,g}\int_0^1\frac{{\rm d} z}{z}\,\frac{\alpha_s}{2\pi}
  \big[P_{ab}(z)\big]_+\,f_{b}\left(\frac{x}{z},t\right)
\end{equation}
with Markovian Monte-Carlo algorithms, i.e.\ by interpreting the evolution 
kernels $P_{ab}$ as probabilities with which to distribute real emissions.
The virtual corrections defined at the kinematic endpoints (defined by the 
$+$ prescription) are included by enforcing probability conservation when 
proposing state changes from parton branching. As a direct consequence, parton
showers commonly generate only those virtual momentum and color exchanges 
that can be obtained from real-emission configurations upon 
integration\footnote{Developments in \cite{Nagy:2016pwq,*Nagy:2017dxh} discuss the extension of 
parton showers to include virtual corrections without direct real-virtual
correspondence.}. 

Modern parton showers 
extend this probabilistic evolution picture to soft radiation by including coherent
emissions from color dipoles. For the simplest case of a color-anticolor
connection, the soft limit of single-gluon emission can be obtained by ensuring
that the sum of evolution kernels recovers eikonal factors, either differentially
or at the integrated level, and by using
the $N_C=3$ value of $C_F=\frac{N_C^2-1}{2N_C}=\frac{4}{3}$.
Parton showers that recover eikonal factors differentially are typically based
on a dipole picture~\cite{Azimov:1984np,*Azimov:1986sf,*Gustafson:1987rq,*Lonnblad:1992tz}. The combined soft-collinear 
evolution of dipole showers is governed by new dipole evolution kernels. These can be obtained by 
the matrix element factorization formula in the soft and collinear limits~\cite{Catani:1996vz}
\begin{equation}
    \label{eq:CS}
    |\mathcal{M}_{n+1}|^2 \simeq -\sum_{\tilde{ij},\tilde{k} \neq \tilde{ij}}^{n} \langle\mathcal{M}_{n}|\frac{\mathbf{T}_{\tilde{k}} \cdot \mathbf{T}_{\tilde{ij}} }{\mathbf{T}_{\tilde{ij}}^2} \mathcal{V}_{\tilde{ij},\tilde{k}}|\mathcal{M}_{n}\rangle,
\end{equation}
where parton $\tilde{ij}$ is the parton undergoing the branching, $\tilde{k}$ 
is a spectator, and $\mathbf{T}_{n}$ is the color matrix element for parton $n$,
and the sum over $\tilde{ij}$, $\tilde{k}$ extends to all colored partons.
This is then further simplified in the $N_C\rightarrow\infty$ limit, and by discarding
spin correlations between the hard matrix element and the branching, to
\begin{equation}
    \label{eq:CSLC}
    |\mathcal{M}_{n+1}|^2 \approx \sum_{\tilde{ij},\tilde{k} \in \text{LC}}^{n} \langle\mathcal{M}_{n}|\mathcal{M}_{n}\rangle \mathcal{V}_{\tilde{ij},\tilde{k}}\,.
\end{equation}
where the sum runs only over partons connected in the $N_C\rightarrow\infty$ 
limit. The dipole kernels $\mathcal{V}_{\tilde{ij},\tilde{k}}$ are often
constructed to only contain two-particle poles in the collinear limits. This
article will assume the construction of Catani and Seymour~\cite{Catani:1996vz} in the
reorganization of~\cite{Schumann:2007mg,Hoche:2015sya}, although the exact form 
of the kernels is not of immediate importance for the following.

The combined soft/collinear evolution equation is solved by iteratively 
proposing, accepting or rejecting state changes according to the parton 
branching kernels $\mathcal{V}_{\tilde{ij},\tilde{k}}$. The
resummation of large logarithmic enhancements into Sudakov form factors
(i.e.\ no-emission or ``parton survival" probabilities) with this procedure requires
that state changes are probed in an ordered sequence. The ordering 
variable should isolate the soft and collinear regions in (at least) the 
single-emission phase space, and thus defines how phase space points are sampled and 
the exponentiation properties of the parton shower. The typical choices
of using angle~\cite{Webber:1983if,*Marchesini:1983bm,*Marchesini:1987cf} or 
hardness-related variables like virtuality~\cite{Sjostrand:1985xi,*Nagy:2014nqa} or 
transverse momentum~\cite{Gustafson:1987rq,*Lonnblad:1992tz,*Schumann:2007mg,*Platzer:2009jq,*Giele:2011cb} are 
related to the on-shell propagator singularities induced by emissions. 
Transverse momenta measured with respect to the axes of a color dipole
have favorable qualities in the soft limit~\cite{Gustafson:1987rq,Dulat:2018vuy} and to define consistent loop 
integration boundaries~\cite{Angeles-Martinez:2015rna}. This article uses the soft transverse
momentum-ordered final-state shower as outlined in~\cite{Hoche:2015sya} and implemented 
in \textsc{Python}  in~\cite{cteq:2015xx} as starting point.

Hardness-ordered parton showers also allow the use of a single scale value to
both avoid the Landau pole in $\alpha_s$ and to transition to 
phenomenological non-perturbative hadronization models. Such hadronization 
models typically employ a leading color QCD picture to assign the starting
conditions for the parton-to-hadron conversion process. For a simple 
infrared-safe treatment of soft and collinear gluons in setting up the
starting conditions for hadronization, we will use the 
leading color Lund string model~\cite{Andersson:1983ia} as implemented in the \textsc{Pythia}~8
event generator~\cite{Sjostrand:2006za,Sjostrand:2014zea}. The subtleties related to this phenomenological step are 
discussed in Sec.~\ref{sec:matching_to_lc}.

The algorithm described in this article remains within the paradigm of 
(weighted) unitary parton shower evolution~\cite{Hoeche:2009xc,Platzer:2011dq,Lonnblad:2012hz}. Thus, the algorithm does not discuss
the inclusion of Glauber gluons~\cite{Collins:1981ta,Bodwin:1981fv,Angeles-Martinez:2015rna,Angeles-Martinez:2016dph,Martinez:2018ffw}, that decouple the color structure of 
virtual corrections from that of real-emission diagrams. It is not obvious that 
the impact of Glauber gluons can be assessed without also including other 
universal virtual corrections, e.g.\ related to analytic continuation or 
remainders of $d$-dimensional loop integration in certain regularization 
schemes. It can be conjectured that such purely virtual corrections could
be included by a ``swing" mechanism~\cite{Lonnblad:1995yk,*Avsar:2006jy,
*Bierlich:2014xba}. Due to these difficulties, we 
postpone the inclusion of Glauber gluons to a future publication.

\section{Parton shower algorithm at fixed $N_C$}\label{sec:Algorithm}

\noindent
This section describes a stochastic algorithm to systematically sample subleading
color configurations in QCD. This relies on standard parton shower
techniques, and will thus be referred to as fixed color (FC) parton shower.
The main goal is then to ensure the color correlations and
interferences between different color structures encoded in Eq.~\ref{eq:CS} are
retained by the FC parton shower algorithm. This idea has previously also been
advocated in~\cite{Platzer:2012np}. Here, we provide an algorithm that allows to handle
an arbitrary number of emissions with less than factorial growth in complexity.
This is realized by directly sampling color configurations in the color flow 
basis~\cite{Maltoni:2002mq} by introducing color-generator-specific dipole splitting 
kernels, which allow to evolve from one definite color configuration to another
definite color configuration. The full color space is sampled stochastically.
This section first reviews the relevant benefits of splitting functions in the color flow basis
(Sec.~\ref{subsec:colorflow}), before discussing the implementation as a FC parton 
shower (Secs.~\ref{subsec:alg} and~\ref{subsec:color_sampling}), and describing
the interface to leading color evolution and hadronization in Sec.~\ref{sec:matching_to_lc}.

\subsection{Splitting Kernels in the Color Flow Basis}\label{subsec:colorflow}

\noindent
The gluon propagator in QCD is proportional to
\begin{equation}
    \langle\left(\mathcal{A}_\mu\right)^{i_1}_{j_1}\left(\mathcal{A}_\mu\right)^{i_2}_{j_2}\rangle \propto \delta^{i_1}_{j_2}\delta^{i_2}_{j_1} - \frac{1}{N_C}\delta^{i_1}_{j_1}\delta^{i_2}_{j_2}.
\end{equation}
Thus, the gluon can thus be treated as two distinct contributions, the 
``nonet" gluon (the terms proportional to $\delta^{i_1}_{j_2}\delta^{i_2}_{j_1}$)
and the ``singlet" gluon (the terms proportional to 
$\delta^{i_1}_{j_1}\delta^{i_2}_{j_2}$). This interpretation is the backbone
of the color flow basis~\cite{Maltoni:2002mq}. Note that the singlet contribution
to the gluon propagator is suppressed by a factor of $1/N_C$. 

The color flow basis can conveniently be represented by a set of basis tensors
represented with color indices ($c_i$) and anticolor 
indices ($\overline{c}_i$)~\cite{Martinez:2018ffw}. In this representation, the
tensors are represented by a color-anticolor pair, which denotes the flow of
color from one leg to another, and can be expressed in terms of 
Kronecker $\delta$'s,
\begin{equation}
    \label{eq:ColorTensor}
|\sigma\rangle = \bigg|\begin{array}{c c c c}1 & 2 & \ldots & n \\ \overline{\sigma}\left(1\right) & \overline{\sigma}\left(2\right) & \ldots & \overline{\sigma}\left(n\right)\end{array}\bigg\rangle = \delta^{c_1}_{\overline{c}_{\sigma\left(1\right)}}\delta^{c_2}_{\overline{c}_{\sigma\left(2\right)}}\ldots\ \delta^{c_n}_{\overline{c}_{\sigma\left(n\right)}},
\end{equation}
where $c_i$ is the color of the $i$th leg and $\overline{c}_{\sigma\left(i\right)}$
is the anticolor of the line connected to the $i$th leg. The color indices 
$c_i$ and $\overline{c}_{\sigma\left(i\right)}$ take on values between 
$1$ and $N_C = 3$. In the color flow basis, the color operators 
($\mathbf{T}$) can be decomposed  as
\begin{equation}
    \label{eq:ColorOperator}
    \mathbf{T}_i = \lambda_i \mathbf{t}_{c_i} - \overline{\lambda}_i \overline{\mathbf{t}}_{\overline{c}_i} - \frac{1}{N}\left(\lambda_i-\overline{\lambda}_i\right) \mathbf{s},
\end{equation}
where $\lambda_i$ and $\overline{\lambda}_i$ are variables used to define the 
type of color object, and their values are given in 
Table~\ref{tab:ColorOperators} for quarks, antiquarks, and 
gluons~\footnote{A similar decomposition in the color-trace basis is discussed in 
Ref.~\cite{Nagy:2007ty}.}. The operators $\mathbf{t}$, $\overline{\mathbf{t}}$, 
and $\mathbf{s}$ are defined by their operation on a basis 
tensor. 

\begin{table}
    \label{tab:ColorOperators}
    \begin{tabular}{| c | c | c |}
    \hline
       & $\lambda_i$  & $\overline{\lambda}_i$ \\
    \hline
    Quark      & $\sqrt{T_R}$ & 0 \\
    Antiquark &    0         & $\sqrt{T_R}$ \\
    Gluon      & $\sqrt{T_R}$ & $\sqrt{T_R}$ \\
    \hline
    \end{tabular}
    \caption{Color variables in the color operators for quarks, antiquarks, and gluons.}
\end{table}

The color flow basis is over-complete and not orthogonal. Therefore, when 
calculating the total color factor for a given amplitude, one needs to sum over
all combinations of products of the basis tensors. For any given amplitude, the
number of possible color flows is given by $(n_{q}+n_g)!\,$, where $n_q$ is the
number of external quark-antiquark pairs, and $n_g$ is the number of external 
gluons. At first glance, this scaling suggests that the color flow basis is
suboptimal to calculate color coefficients for many-parton states. This 
scaling is however outweighed by the fact that the color flow basis allows one to 
efficiently ``build up" the color coefficient for a definite color structure 
with $n$ partons from the coefficients of definite $(n-1)$ parton color 
configurations\footnote{Using a similar method for orthogonal color 
bases~\cite{Keppeler:2012ih} appears to be less straightforward.}.

In order to use the color flow representation in a parton shower, the 
splitting kernels need to be recalculated in the color flow basis. The 
most straight forward case is the splitting of the nonet or singlet gluon into
a quark and an antiquark pair. The color flow for this splitting is 
completely determined by color conservation, and is therefore identical to the 
traditional splitting kernel. Gluon emission from (anti)quarks in the FC
parton shower should be governed by two splitting kernels -- one to describe
the emission of a nonet gluon ($P^{(9)}_{qq}$), and another for singlet gluon 
emissions ($P^{(1)}_{qq}$). Using Catani-Seymour dipole 
kernels~\cite{Catani:1996vz} in the variables of the \textsc{Dire} parton 
shower~\cite{Hoche:2015sya}
\begin{align}\label{eq:LCAcceptReject}
P_{qq} = C_F\left(\frac{2\left(1-z\right)}{\left(1-z\right)^2+\kappa^2}-\left(1+z\right)\right),\quad
P_{gg} = \frac{C_A}{2}\left(\frac{2\left(1-z\right)}{\left(1-z\right)^2+\kappa^2}-2+z\left(1+z\right)\right), \quad
P_{gq} = \frac{T_R}{2}\left(1-2z\left(1-z\right)\right) \nonumber
\end{align}
to capture the phase space dependence of the branching, the
color flow specific kernels can be defined as
\begin{eqnarray}
    P^{(9)}_{qq}(x)
 &=& \lambda_i\mathbf{t}_{c_{i}} \frac{P_{qq}(x)}{C_F}\\
    P^{(1)}_{qq}(x)
 &=& -\frac{\lambda_i}{N_C}\mathbf{s}\frac{P_{qq}(x)}{C_F}\, .
\end{eqnarray}
One splitting kernel is included for each possible change (due to the
action of $\mathbf{t}_{c_{i}}$) to the color structure, with
some examples given in Fig~\ref{fig:ColorFactorExample}.
The branching of a gluon into two gluons can be distributed over two nonet 
gluon splitting kernels,
\begin{align}
    P^{(+)}_{gg} &= \lambda_i\mathbf{t}_{c_{i}}\frac{P_{gg}(x)}{C_A},\\
    P^{(-)}_{gg} &= -\overline{\lambda}_i\overline{\mathbf{t}}_{\overline{c}_{i}}\frac{P_{gg}(x)}{C_A},
\end{align}
where $P^{(+)}_{gg} (P^{(-)}_{gg})$ is the splitting off of the 
color(anticolor) gluon line, respectively. Examples are again shown
in Fig~\ref{fig:ColorFactorExample}. Note that in the color flow 
basis, nonet and singlet gluons do not couple directly. Hence, splitting 
functions involving both nonet and singlet gluons are absent.

\subsection{Implementation as a Parton Shower}\label{subsec:alg}

\noindent
The leading color approximation usually employed in parton showers removes 
all color interference effects at $\order{1/N_C^2}$. Naively, this would also
occur when using the color-flow specific splitting functions described in the
previous section, although some subleading contributions due to extended
spectator assignments and the presence of singlet gluons would be retained.

\begin{figure}
    \label{fig:bra-ket}
    \includegraphics[angle=90,width=0.5\textwidth]{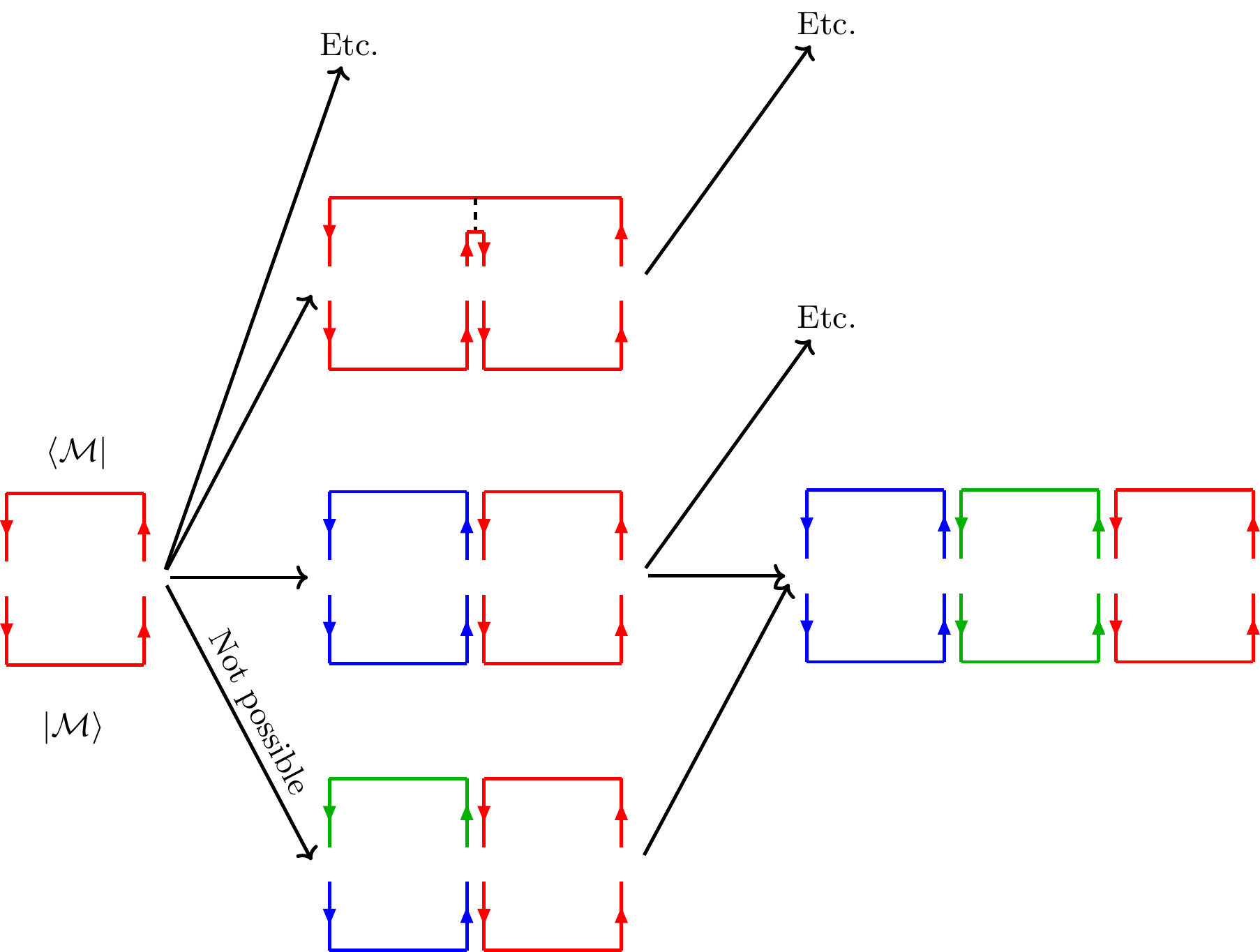}
    \caption{Example for the sampling of color configurations. Note that both
    color configurations $\langle \mathcal{M} |$ and $| \mathcal{M} \rangle$ 
    are kept throughout the evolution. A detailed discussion can be found in the
    main text.}
\end{figure}

The recovery of interference terms is central to defining a consistent FC
parton shower. One possible solution is to sample color configurations 
stochastically, i.e.\ organize the evolution of the color configuration as 
a step-by-step splitting process from one definite color structure to another
definite color structure. The correct inclusion of color factors then requires
that there exists at least one path of splittings populating a definite
color configuration, and that the correct color-factor of the
configuration can be calculated efficiently. This strategy is illustrated in
Fig.~\ref{fig:bra-ket}. There, the $\langle\mathcal{M}|$ state contains the
emission from the spectator line, while the $|\mathcal{M}\rangle $ state 
includes the emission from the radiator line. In leading color showers, both 
amplitudes are identical. The first step in moving to FC showers, is to remove 
this simplification. This is easily achieved for the first emission from a
single color-anticolor dipole (i.e.\ with only one possible assignment of
radiator and spectator) by employing an eikonal radiation pattern.

As already alluded to, complications arise in handling the 
states $\langle\mathcal{M}'|^* \neq |\mathcal{M}\rangle$ after the first 
emission: to recover Eq.~\ref{eq:CS}, the spectator assignment needs to be 
extended to include all partons. Interference effects can be handled by a Monte
Carlo sum over states. For this, both  $\langle\mathcal{M}'|$ and 
$|\mathcal{M}\rangle$ are kept throughout the evolution. For a change 
$|\mathcal{M}\rangle\rightarrow|\mathcal{\tilde{M}}\rangle$ that was chosen with the 
Sudakov veto algorithm, a definite $\langle\mathcal{\tilde{M}}'|$ is chosen based 
on all possible non-zero probabilities 
$\langle\mathcal{\tilde{M}}'| \mathcal{\tilde{M}}\rangle$. Both $\langle\mathcal{\tilde{M}}|$
and $\mathcal{\tilde{M}}'\rangle$ are then used in the next evolution step when 
probing a subsequent color configuration. This is permissible if we choose to keep the 
phase-space dependence of the splitting kernels independent of the color configuration 
$\langle\mathcal{M}'|$. 

A technical complication that arises when adding in the interference effects
in this Monte-Carlo fashion is that emissions from the radiator and emissions 
from the spectator are not necessarily associated with the same momentum
assignments. This necessarily deteriorates the efficiency of the Monte-Carlo
integration over phase space, since the phase-space structure probed by 
emissions from the radiator might approach different singular phase space
points than the structure probed by emissions from the spectator in the 
$k_T\to 0$ limit. To counter this issue, a cutoff $t^{\text{cut}}_{FC}$ to transition from a 
FC shower to an LC shower is implemented.
If the algorithm decides to admit an emission below this 
cutoff, then only the leading color connected parton can be the spectator, and
the LC color factor is used. While this does introduce some error in the
calculation, the cutoff can be pushed sufficiently low to have a negligible
effect on the distributions as will be shown in Section~\ref{sec:results}. The
main benefit of this cutoff is that the weight fluctuations are reduced and the
convergence of the calculation is improved, requiring fewer events and less time to 
obtain smooth distributions.

With the issues discussed above in mind, the FC parton shower algorithm is 
described in detail below. The algorithm uses two accept-reject steps, along 
with a weighting factor to correct for the sign of the color factor. 
The first step in the algorithm involves a modification to the allowed colors 
for partons. Typically, for a leading color shower, the partons are labeled 
with a color and an anticolor index ranging from 1 to $\infty$. In the FC 
parton shower, color and anticolor indices range from 1 to 3 (i.e.\ $r$, $g$, 
and $b$). This begins with the labeling of the partons that are produced in the
hard process. Multiple partons can have the same color index. This is important 
when calculating the color factor, and allows the calculation to be performed 
more efficiently.

The first accept-reject step is similar to a leading color shower, but with the
extension that non-leading color connected partons are allowed to be 
spectators. 
This can be viewed as determining if a splitting is kinematically 
preferred: the accept-reject step determines the phase space variables $(t,z)$, 
which parton is considered the radiator and which the spectator.
The phase space region $t < t^{\text{cut}}_{FC}$ will be considered at leading color
only. In this case, no further steps related to picking the color structure
are necessary. Splittings with $t > t^{\text{cut}}_{FC}$ will be considered
with complete $N_C=3$ color factors, and the color factor needs to be corrected
to obtain the full color result, if the splitting is not a gluon into a 
quark-antiquark pair. This is facilitated by calculating the color
correlator
\begin{equation}
    \langle \mathcal{M}'| \frac{\mathbf{T}_{\tilde{k}} \mathbf{T}_{\tilde{ij}}}{\mathbf{T}_{\tilde{ij}}^2} | \mathcal{M} \rangle,
\end{equation}
where the $\mathbf{T}$'s are given by Eq.~\ref{eq:ColorOperator}, and depends 
on the type of partons $k$ and $ij$. This factor is calculated exactly, given 
a fixed radiator and spectator. This result is then used as a second 
accept-reject probability to correct to splitting kernel from the LC overestimate 
(produced by the first accept-reject step) to the correct overall color factor 
for the given radiator and spectator. Since the accept probability is 
not bounded by zero and one, we employ weighted parton shower 
techniques~\cite{Hoeche:2009xc,Platzer:2011dq,Lonnblad:2012hz} in
the second accept-reject step to ensure that the correct overall color factor is
consistently exponentiated\footnote{Specifically, splittings will be accepted 
with probability $f/g$, and introduce corrective event
weights. If the event is accepted, it receives the weight $\frac{g}{h}$. 
Rejection leads to an event weight $\frac{g}{h}\frac{h-f}{g-f}$. Here,
$f$, $g$, and $h$ are given by
\begin{align*}
    f = -\langle\mathcal{M}'|\mathbf{T}_{k}\cdot\mathbf{T}_{ij}|\mathcal{M}\rangle\, ,\qquad
    h = \langle\mathcal{M}'|\mathbf{T}_{ij}^2|\mathcal{M}\rangle\, ,\qquad
    g = 
    \begin{cases}
    -h & \frac{f}{g} < 0 \\
    2 f & \frac{f}{g} > 1 \\
    h & \text{otherwise}
    \end{cases}.
\end{align*}}. At this point, we have achieved the exponentiation of 
Eq~\ref{eq:CS}. To allow iteration, it is necessary to fix a definite color
structure for subsequent evolution. This will be discussed in the next section.

\subsection{Color Flow Sampling}\label{subsec:color_sampling}

\begin{figure}
    \label{fig:ColorFactorExample}
    \subfigure[$P^{(9)}_{rr}$]{
        \includegraphics[width=0.2\textwidth]{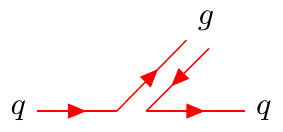}
    }
    \subfigure[$P^{(9)}_{rg}$]{
        \includegraphics[width=0.2\textwidth]{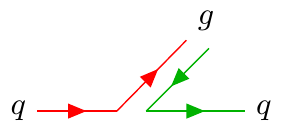}
    }
    \subfigure[$P^{(9)}_{rb}$]{
        \includegraphics[width=0.2\textwidth]{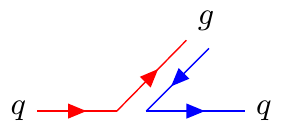}
    }\\\vspace*{0.3cm}
    \subfigure[$P^{(1)}_{rr}$]{
        \includegraphics[width=0.2\textwidth]{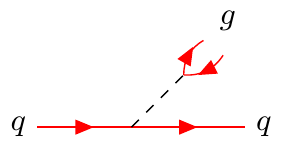}
    }
    \subfigure[$P^{(1)}_{rg}$]{
        \includegraphics[width=0.2\textwidth]{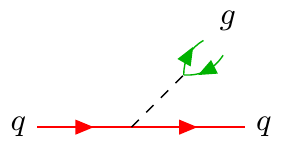}
    }
    \subfigure[$P^{(1)}_{rb}$]{
        \includegraphics[width=0.2\textwidth]{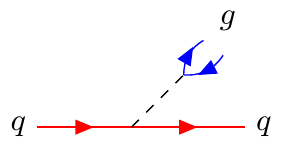}
    }\\\vspace*{0.3cm}
    \subfigure[$P^{(+)}_{rr}$]{
        \includegraphics[width=0.2\textwidth]{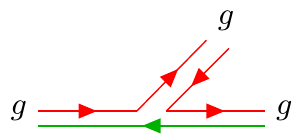}
    }
    \subfigure[$P^{(+)}_{rg}$]{
        \includegraphics[width=0.2\textwidth]{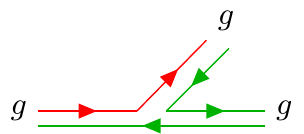}
    }
    \subfigure[$P^{(+)}_{rb}$]{
        \includegraphics[width=0.2\textwidth]{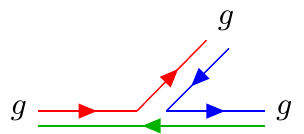}
    }\\\vspace*{0.3cm}
    \subfigure[$P^{(-)}_{gr}$]{
        \includegraphics[width=0.2\textwidth]{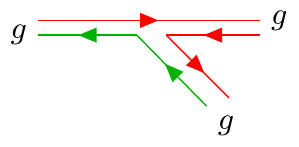}
    }
    \subfigure[$P^{(-)}_{gg}$]{
        \includegraphics[width=0.2\textwidth]{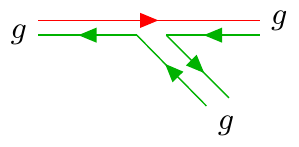}
    }
    \subfigure[$P^{(-)}_{gb}$]{
        \includegraphics[width=0.2\textwidth]{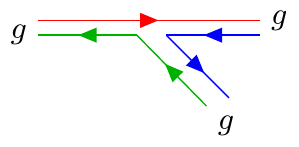}
    }
    \caption{Examples of possible splittings in the fixed color 
    shower. The upper two rows show all possible splittings for a red quark,
    while the lower two rows show all splittings of a red-antigreen gluon. Note
    that in all cases, we implement one kernel for soft-tagged emission and
    one kernel for soft-tagged radiator.}
\end{figure}

\noindent
The overall color factor exponentiated into no-emission probabilities now 
correctly recovers the full color correlator (Eq.~\ref{eq:CS}). To allow the 
iteration of the stochastic color sampling algorithm once a branching
has been accepted, it is necessary to choose a definite color structure
for the branching. This choice has to be commensurate with the contribution
of the particular color configuration to the overall color factor.
To obtain the correct result in the color flow basis, the color operators can 
be separated into different components, depending on if the splitting is 
$P^{(9)}_{qq}$, $P^{(1)}_{qq}$, $P^{(9)}_{\overline{q}\overline{q}}$, 
$P^{(1)}_{\overline{q}\overline{q}}$, $P^{(+)}_{gg}$, or $P^{(-)}_{gg}$. 
Six possible color structures for the radiator and six possible color 
structures for the spectator are allowed for given emitter $ij$ with color 
$c_{i}$ and spectator $k$ with anticolor $\overline{c}_{k}$, giving in total
36 possible configurations. Possible splittings of a red quark and a red-antigreen gluon are illustrated
in Fig.~\ref{fig:ColorFactorExample}. An additional event weight needs to be 
applied to correctly account for picking one definite (out of 36 possible)
color structures.

The correction is obtained by first calculating the ``color weight" for each one of the 
36 possible configurations,
\begin{equation}
    P_{\alpha\beta}
 = |\langle\mathcal{M'}|\mathbf{t}^{\alpha}_{k}\cdot\mathbf{t}^{\beta}_{ij}|\mathcal{M}\rangle|\, ,
\end{equation}
where $\mathbf{t}^{\alpha}_{k}$ is one of the six possible color factors for 
the color insertion on the spectator line acting on $\langle\mathcal{M'}|$, 
and $\mathbf{t}^{\beta}_{ij}$ is color insertion on the radiator line acting 
on $|\mathcal{M}\rangle$. The absolute value ensures a positive 
definite probability to pick one color structure,
\begin{equation}
    \label{eq:color_probablility}
    P
 = \frac{P_{\alpha\beta}}{\sum_{\alpha,\beta} P_{\alpha\beta}}\, .
\end{equation}
The color structure can then be chosen probabilistically according to
Eq.~\ref{eq:color_probablility}. Finally, to obtain the correct weight for 
the branching, the sign of this contribution to the overall color factor needs
to be reinstated. Hence, an additional event weight 
\begin{equation}
\frac{g_{col}}{h_{col}} = 
    \frac{\mathbf{t}^{\alpha}_{k}\cdot \mathbf{t}^{\beta}_{ij}}
         {|\mathbf{t}^{\alpha}_{k}\cdot \mathbf{t}^{\beta}_{ij}|}
    \frac{\sum_{\alpha,\beta}|\mathbf{t}^{\alpha}_{k}\cdot \mathbf{t}^{\beta}_{ij}|}
         {\sum_{\alpha,\beta}\mathbf{t}^{\alpha}_{k}\cdot \mathbf{t}^{\beta}_{ij}}.
\end{equation}
is applied. At this point, the emission is generated according to 
Eq.~\ref{eq:CS}, and a definite color configuration is chosen to allow
iteration of the stochastic color sampling. The real-emission kinematics are
constructed and, and the algorithm returns to the beginning again. 

If a gluon decays into a quark-antiquark pair, the color factor is determined 
explicitly by color conservation. Hence, once a proposed branching passes
the first accept-reject step, the kinematics
of the quark and antiquark are generated and the next emission is considered.

The complete algorithm is summarized, for convenience, as a flow chart
in Fig.~\ref{fig:AlgorithmFlowchart}. An example of the first two branchings 
for the evolution of $\langle \mathcal{M} |$ and $|\mathcal{M}\rangle$ is shown 
in Fig.~\ref{fig:bra-ket}. In this example, we begin with a red quark and an 
antired antiquark, and calculate all 36 possible color configurations after 
one emission. Only three of the 36 are shown explicitly in the example. 
The probabilities of ending up in each of the three sample states shown are,
from the top down, $\frac{1}{12}$, $\frac{1}{4}$ and zero. The last contribution
is vanishing, because the inner product for the given two color flows is 
exactly zero. However, the algorithm still explores all of color space, even if
one history to reach the desired state has zero probability -- as indicated by
the central row. As long as any history with non-vanishing probability
exists, the color configuration is sampled, including the correct color factor.

\subsection{Further technicalities}\label{sec:matching_to_lc}

\noindent
The FC parton shower algorithm relies on weighted parton shower 
techniques to ensure correct exponentiation of the color 
correlator. As such, it is possibly prone to numerical instabilities because
of the accumulation of large event weights if 
$\langle\mathcal{M}'|\mathbf{T}_{ij}^2|\mathcal{M}\rangle$ is a very poor
overestimate of $-\langle\mathcal{M}'|\mathbf{T}_{k}\cdot\mathbf{T}_{ij}|\mathcal{M}\rangle$.
If this is repeatedly the case, event weights can become large, thus requiring 
many more events to appropriately obtain the true result. 
This can for example be the case if high-multiplicity configurations
are probed very often at small ordering parameter. Therefore, to mitigate this 
effect, we introduce a color sampling cutoff scale ($t_{FC}^{\text{cut}}$). Below 
this scale, we match the FC parton shower to a leading color (LC) parton shower. 
Note that we do not restrict the number of emissions that carry the correct
color factor. Matching the FC parton shower to leading color using only a
resolution scale defines an infrared safe method, whereas infrared safety would
have to be re-evaluated if the FC parton shower were terminated after a fixed
number of emissions.

Since both the LC parton shower and the FC parton shower rely on the 
color flow basis, the transition to leading color is straightforward: an auxiliary
color structure with leading color connected partners is tracked during FC 
evolution. The only complication that arises is in handling the color singlets. 
In this case, mapping onto a leading color configuration means that the emitting
parton keeps the original leading color connected partner, and the singlet
is only leading color connected to itself. Additionally, singlet gluons 
in the auxiliary leading color configuration will only be allowed branch into 
quark-antiquark pairs once the LC parton shower evolves the state. This follows
since gluon radiation from the color line is exactly canceled by gluon radiation from the 
anticolor line. It is beneficial to enforce this cancellation exactly, rather than
achieving it stochastically by averaging over events. 

Once the parton shower evolution is terminated at the overall
cutoff $t^{\text{cut}}$, the event transitions to subsequent hadronization.
In this paper, we employ the Lund string model~\cite{Andersson:1983ia} as 
implemented in \textsc{Pythia} 8.2~\cite{Sjostrand:2006za,Sjostrand:2014zea} for 
hadronization. This model is constructed in the 
$N_C\to \infty$ limit, and thus not immediately suitable for projecting $N_C=3$
partonic states onto a spectrum of hadrons. However, note that the 
$N_C \to \infty$ approximation is less ad-hoc since the FC parton shower is not
used all the way down to $t^{\text{cut}} < t_{FC}^{\text{cut}}$, meaning that
a perturbative mapping to a leading color configuration has already been 
obtained. Beyond this mapping, it is important to clarify how $N_C=3$ color 
configurations are prepared for hadronization. This in particular concerns the
treatment of gluons. 

Within the Lund string model, gluons are interpreted as links in a chain of
color dipoles, i.e.\ a color string. These links only induce transverse 
``kinks" on the string, as further parameters would impede an infrared safe
matching onto partonic states. When preparing a fixed color configuration
for hadronization, we need to ensure that this quality is preserved. 
Otherwise, perturbative cancellations between color singlet and nonet gluon
states are distorted and potentially invalidated, and the result is not 
infrared safe. To avoid these issues, a second auxiliary leading color 
configuration is kept throughout the (FC and LC) evolution. In this 
configuration, all emissions of a singlet gluon are replaced by a nonet
gluon, retaining the kinematics of (FC or LC) event. This event is
not used to calculate color factors, and is only used as input for
hadronization, instead of the final event after the parton shower has terminated. 
This guarantees that hadronization effects do not spoil the cancellation
required to remove the trace component of nonet gluons by means of singlet
gluons.

\section{Results}\label{sec:results}

We have as implemented the FC
parton shower as \textsc{Python}  code, building on the implementation in~\cite{cteq:2015xx}. 
The code is interfaced to \textsc{Pythia} through Les Houches event 
files~\cite{Alwall:2006yp}. This section presents results of the new shower 
algorithm for $e^+e^-\rightarrow$ jets scattering events, split into a validation
section and some preliminary comparisons to LEP data. All results have been
generated with an overall parton shower cutoff $t^{\text{cut}} = 1.0$ GeV and 
a fixed color cutoff of $t_{FC}^{\text{cut}} = 3$ GeV.
The LC baseline used as a comparison is a direct implementation of the
final-state shower presented in~\cite{Hoche:2015sya}. This baseline
uses $C_F=\frac{4}{3}$ for gluon emissions from (anti)quark lines, and
explicitly includes two splitting kernels for the two possible
color-assignments for the $g\rightarrow g g$ branching.

\subsection{Validation}

\noindent
This section illustrates the differences of FC and LC parton showers. To gain
insight into these differences, we use observables that should be 
particularly sensitive. The impact of color corrections can be significant
when examining states of fixed parton multiplicity, e.g.\ after the second 
parton shower emission. Instead, the observables used in this section will
be based on jets to ensure a realistic assessment of the impact.

At the level of the second emission, FC and LC parton shower start deviating, 
because color factors become non-trivial, and because non-leading color
radiator-spectator dipoles appear. Color corrections should thus influence the
distribution of four-jet events, in particular the differences of dijet 
invariant masses in four-jet events. Jet masses are heavily affected by
non-perturbative corrections. This suggests testing the impact of color 
corrections on the dijet mass ratio variables
$r_{ij,kl} = \frac{s_{ij}}{s_{kl}}$, where $s_{ij} = (p_i+p_j)^2$ and $p_i$ are 
energy-ordered jet momenta. The two hardest jets (index $3$ and $4$) should be a 
reasonable proxy for regions enriched by the primary quark/antiquark.
Secondary emissions from the quark-antiquark dipole are allowed in the
FC parton shower, but absent in the LC version. We thus expect ratios 
containing $s_{34}$ to exhibit color corrections. A similar argument holds
for the ``eikonal pattern" $\kappa_{364} = \frac{s_{36}s_{64}}{s_{34}s_{3456}}$,
which probes the radiation from the quark-antiquark dipole.

\begin{figure}
    \label{figs:invariants_allemit}
    \subfigure[][]{
        \includegraphics[width=0.3\textwidth]{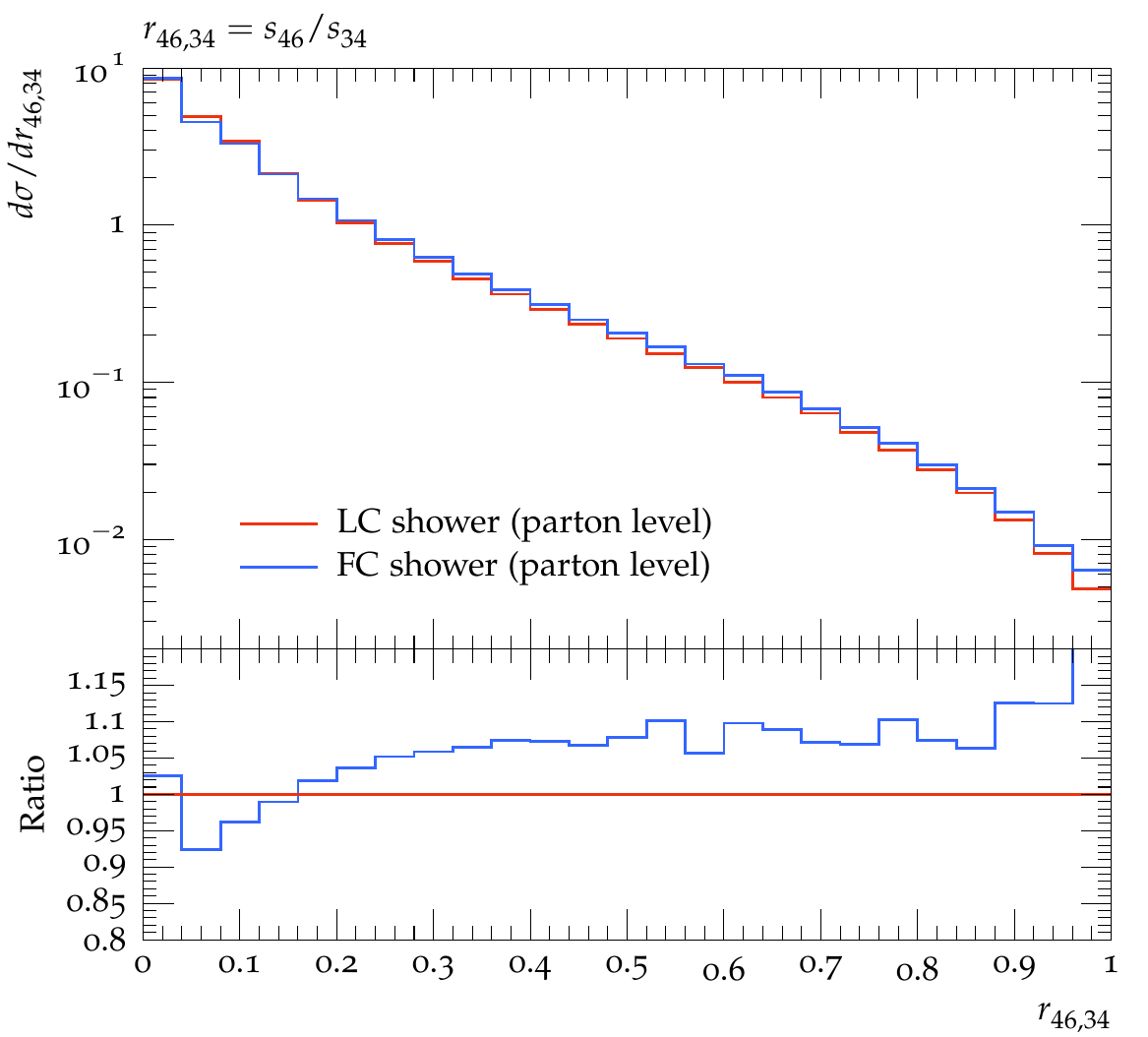}
        \label{figs:nohad:s46_s34}
    }
    \subfigure[][]{
        \includegraphics[width=0.3\textwidth]{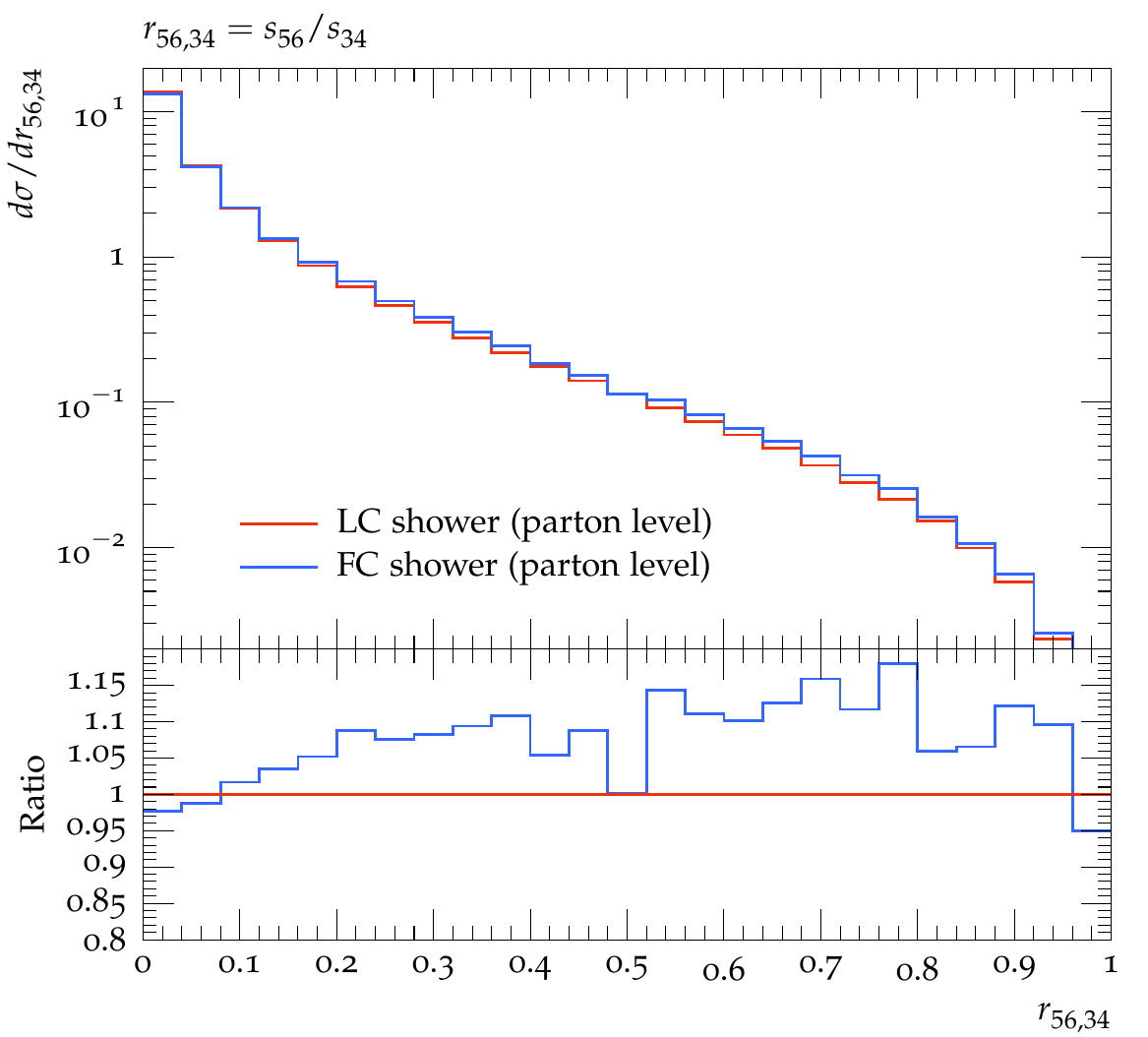}
        \label{figs:nohad:s56_s34}
    }
    \subfigure[][]{
        \includegraphics[width=0.3\textwidth]{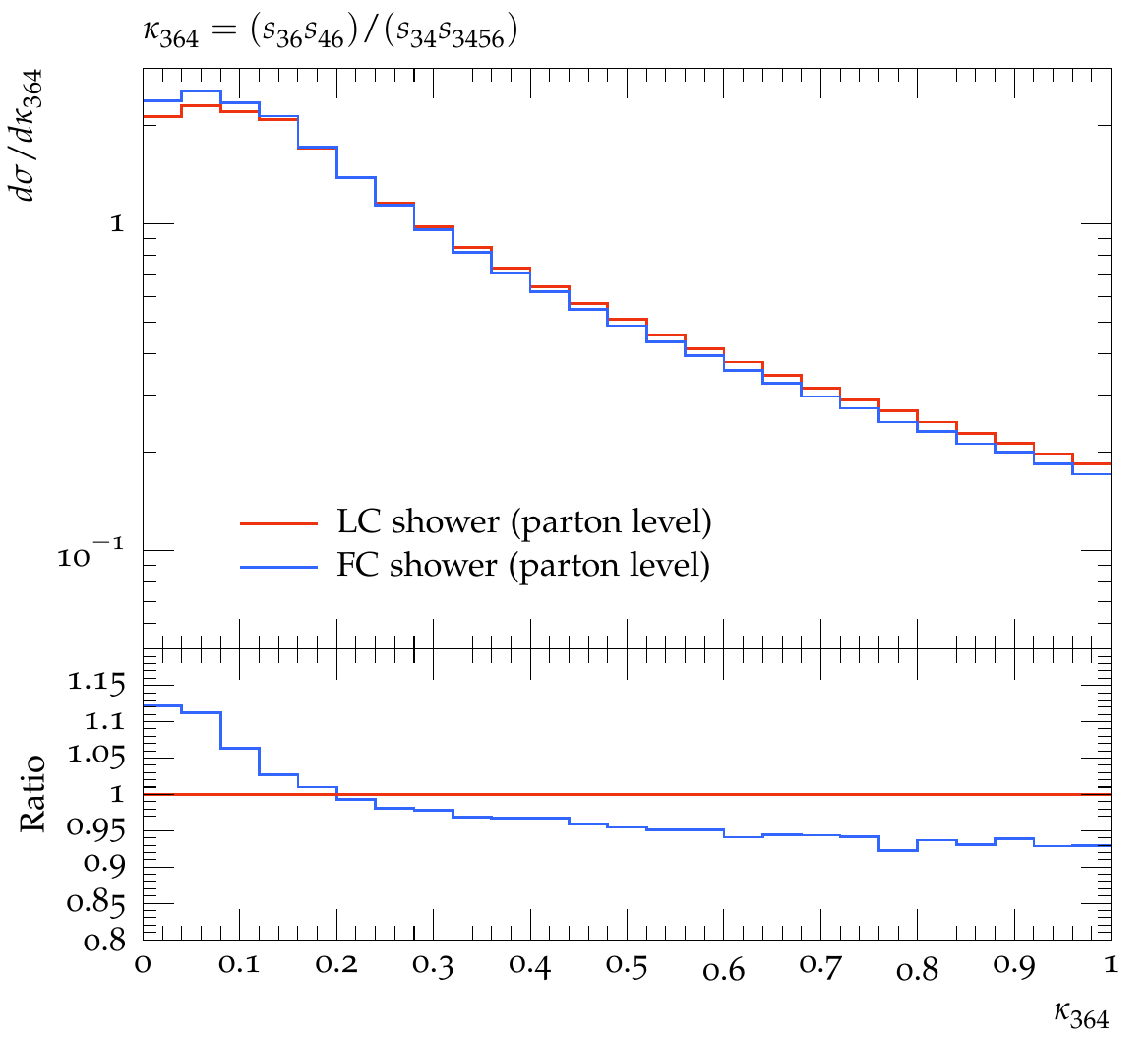}
        \label{figs:nohad:eik}
    }\\
    \subfigure[][]{
        \includegraphics[width=0.3\textwidth]{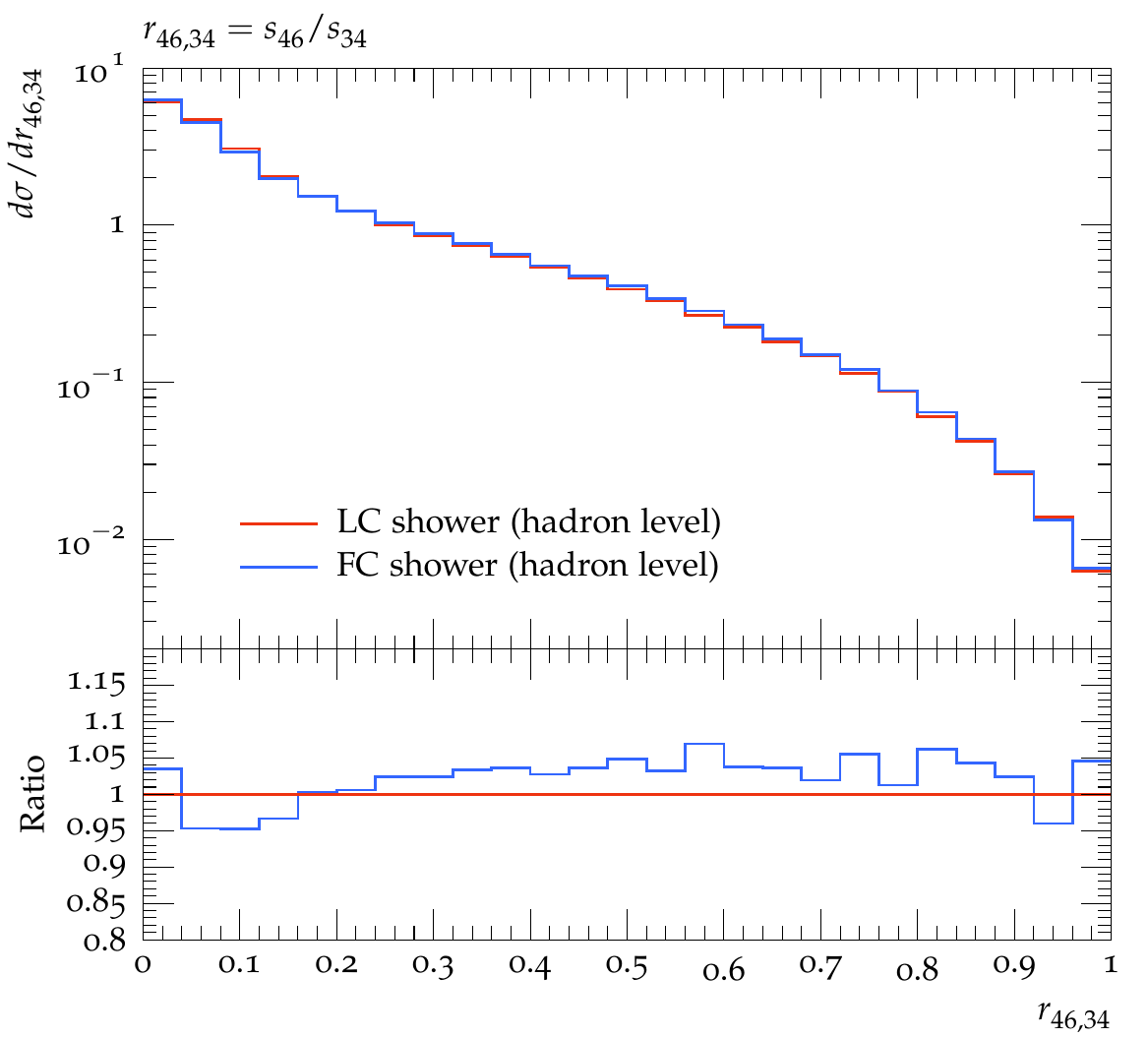}
        \label{figs:had:s46_s34}
    }
    \subfigure[][]{
        \includegraphics[width=0.3\textwidth]{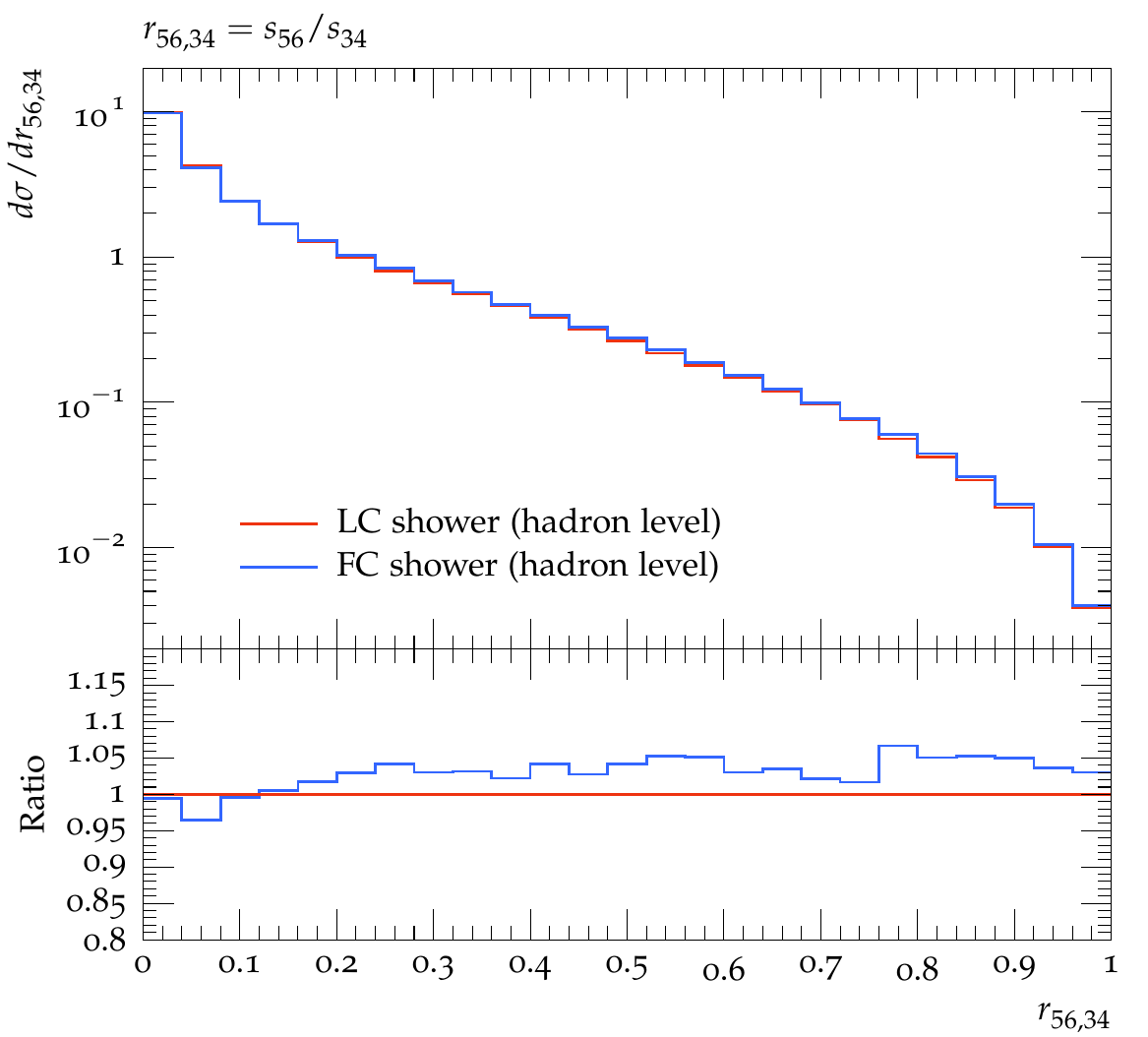}
        \label{figs:had:s56_s34}
    }
    \subfigure[][]{
        \includegraphics[width=0.3\textwidth]{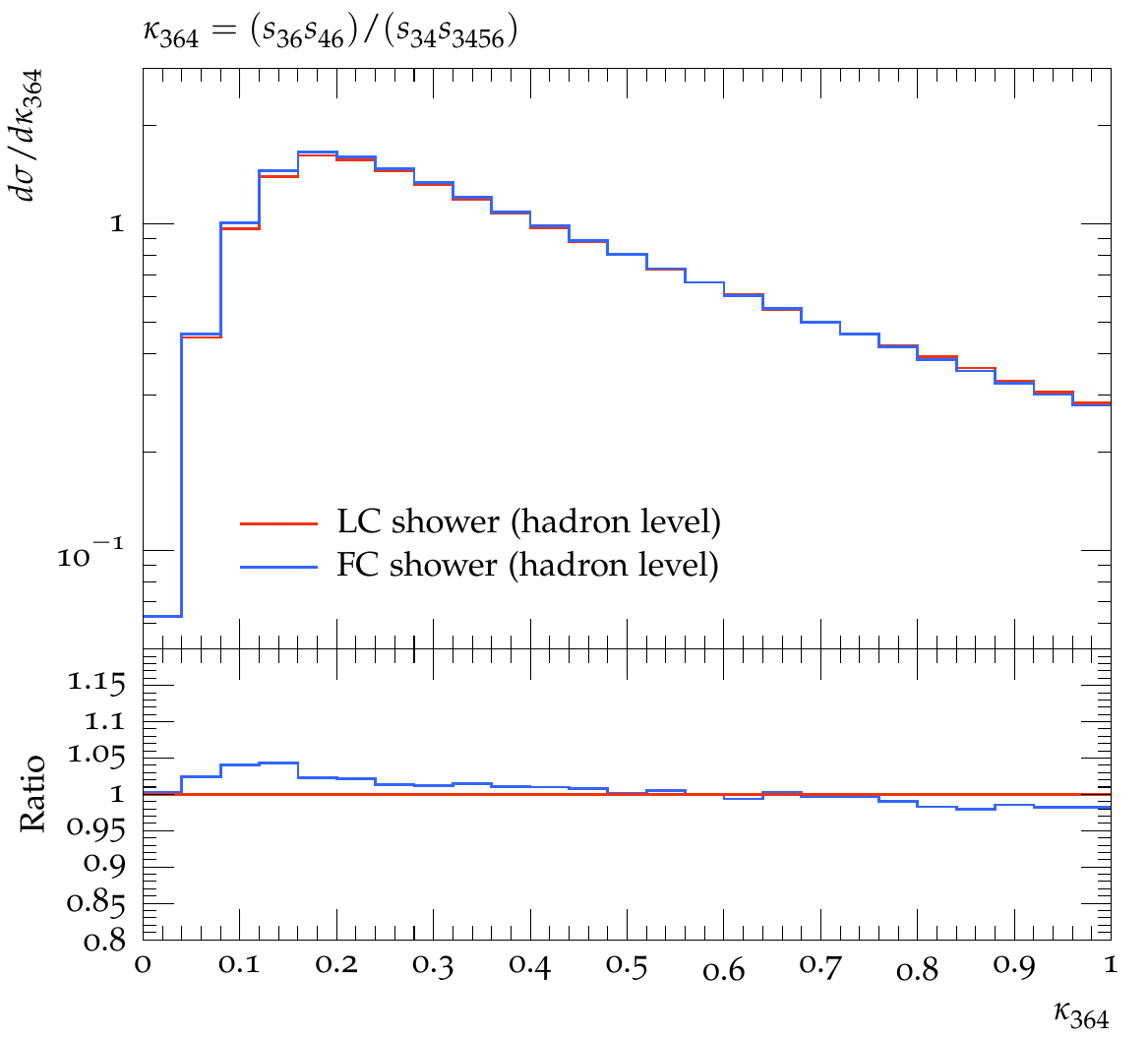}
        \label{figs:had:eik}
    }
    \caption{Comparison of the LC parton shower to the FC parton shower. 
    The definition of the dijet mass ratios $r_{ij,kl}$ and the scaled jet 
    eikonal can be found in the main text. The upper row gives parton level
    results, while the lower row shows results at hadron level. 
    are defined using the Durham algorithm~\cite{Catani:1991hj} as implemented in the 
    \\textsc{Fastjet} package~\cite{Cacciari:2011ma}, clustering to exactly four jets with $p_T>3$~GeV.
    All figures have been produced using RIVET~\cite{Buckley:2010ar}.}
\end{figure}

Representative results are shown in Fig~\ref{figs:invariants_allemit}. As 
expected, the FC parton shower yields a reduced $s_{34}$, thereby producing
harder $r_{46,34}$ and $r_{56,34}$ spectra in Figs.~\ref{figs:nohad:s46_s34}
and \ref{figs:nohad:s56_s34}, respectively. The eikonal pattern $\kappa_{364}$
(Fig.~\ref{figs:nohad:eik}) also shows the impact of the FC parton shower. 
Figures~\ref{figs:had:s46_s34}-\ref{figs:nohad:eik} however accentuate that
the impact of hadronization is still non-negligible. At hadron level, the 
color corrections of $\mathcal{O}(5\%)$ are found.

\subsection{Comparison to data}

\noindent
This section compares the LC and FC parton shower to selected LEP data. 
Since the FC parton shower directly builds upon a \textsc{Python}  implementation 
for Dire, we use the default tune for the \textsc{Dire} plugin to \textsc{Pythia} 
(see App.~\ref{app:tune}) to set non-perturbative parameters. This tune
was performed with a version of \textsc{Dire} that includes $\mathcal{O}(\alpha_s^2)$
corrections designed to recover the inclusive NLO corrections to soft-gluon
emissions, i.e.\ the two-loop cusp anomalous 
dimension~\cite{Kodaira:1981nh,Davies:1984hs,Davies:1984sp,Catani:1988vd}.
In the current study, we do not include these corrections, since the two 
loop cusp anomalous dimension partially arises from sub-leading color effects
which are already captured by the FC parton shower. To avoid double-counting,
and to keep the comparison on even footing, we therefore do not include such 
corrections in the (LC or FC) parton shower. This deteriorates the comparison
to LEP data, so we use $\alpha_s\left(M_Z\right)=0.125$ for a more sensible
baseline. This discussion already emphasizes that the quality of the data 
comparisons should not be considered final before the event generator has been
retuned. Such a procedure will however obscure the differences between LC and FC
shower, and is thus avoided here.

\begin{figure}
    \begin{minipage}[b]{0.45\textwidth}
    \includegraphics[width=\textwidth]{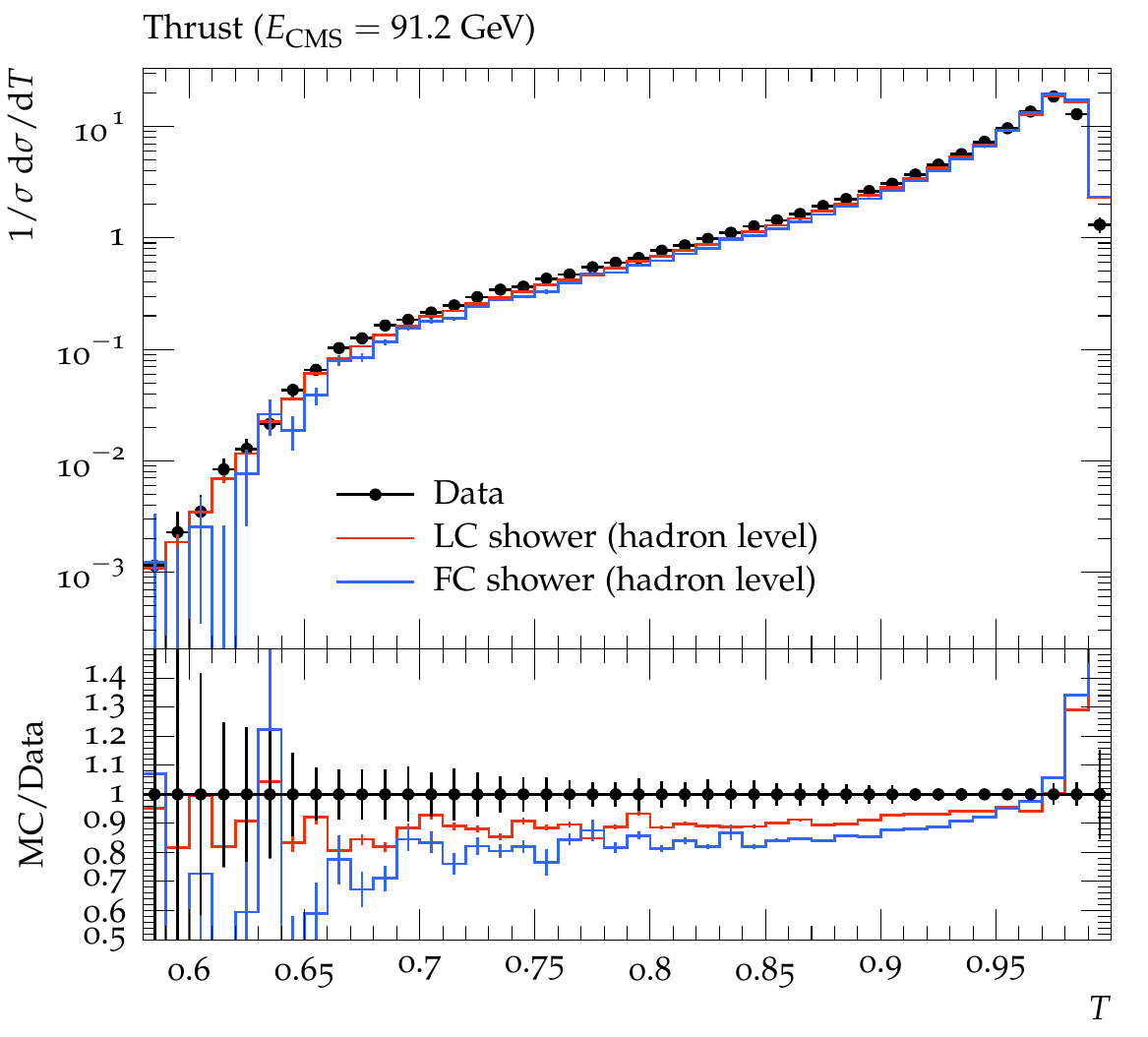}
    \caption{Thrust as measured by ALEPH~\cite{Heister:2003aj} and 
    implemented in RIVET~\cite{Buckley:2010ar}, compared to leading color and fixed color parton showers.}
    \label{fig:Thrust}
    \end{minipage}
    \hfill
    \begin{minipage}[b]{0.45\textwidth}
    \includegraphics[width=\textwidth]{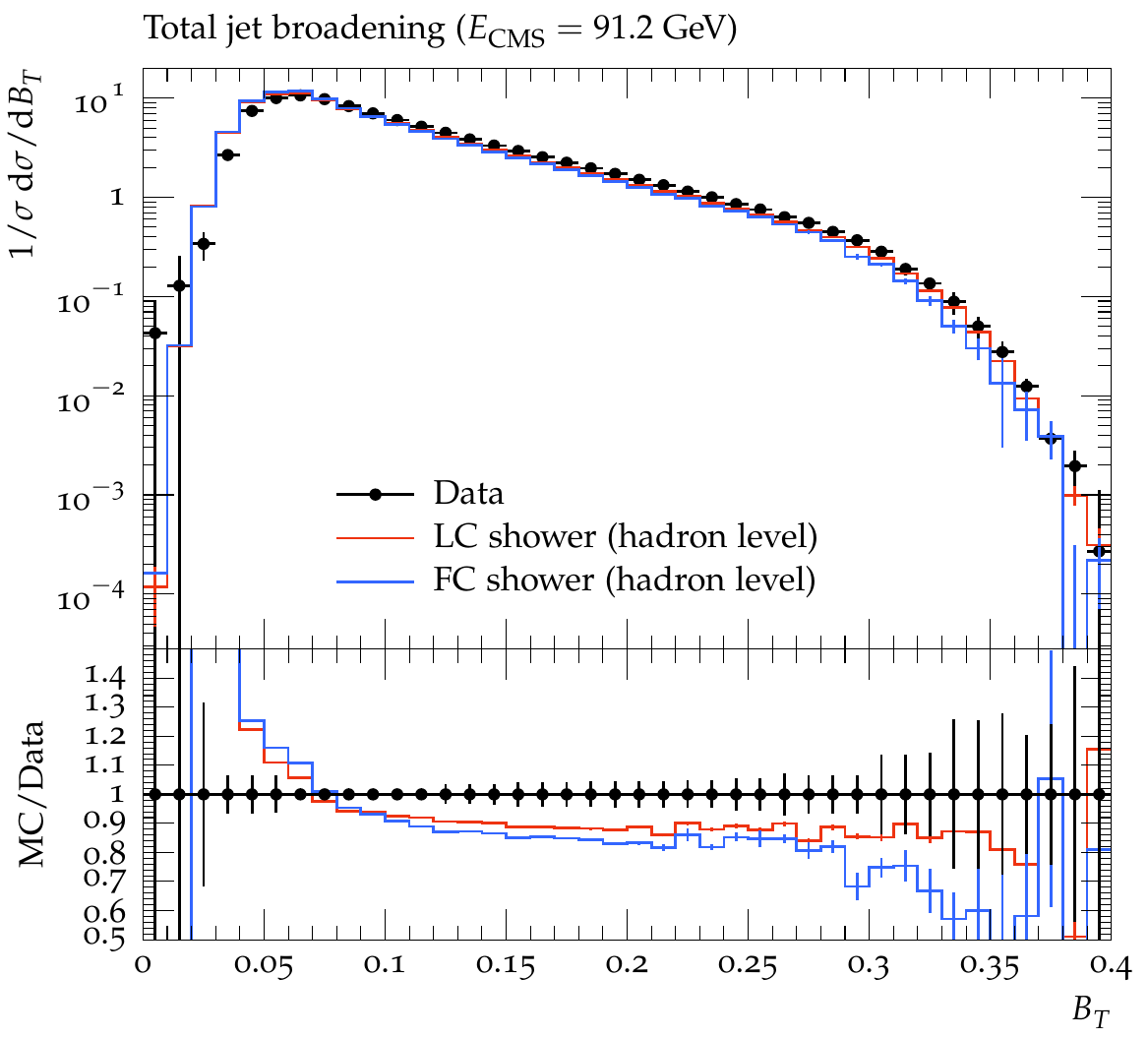}
    \caption{Total jet broadening as measured by ALEPH~\cite{Heister:2003aj} and 
    implemented in RIVET~\cite{Buckley:2010ar}, compared to leading color and fixed color parton showers.}
    \label{fig:JetBroadening}
    \end{minipage}
\end{figure}

Figure~\ref{fig:Thrust} compares the LC and FC parton showers
to ALEPH data~\cite{Heister:2003aj}. As anticipated, the parton showers are not an ideal
representation of the data. The FC shows a reduced rate of semi-hard emissions
compared to LC, which also induces a reduced Sudakov suppression at small 
thrust. A similar trend can also be observed in the related jet broadening
shown in Fig.~\ref{fig:JetBroadening}. This reduction of overall activity
can be traced to negative interference contributions due to the presence
of color singlet gluons. The overall effect of subleading contributions is,
however, moderate.

\begin{figure}
    \begin{minipage}[b]{0.45\textwidth}
    \includegraphics[width=\textwidth]{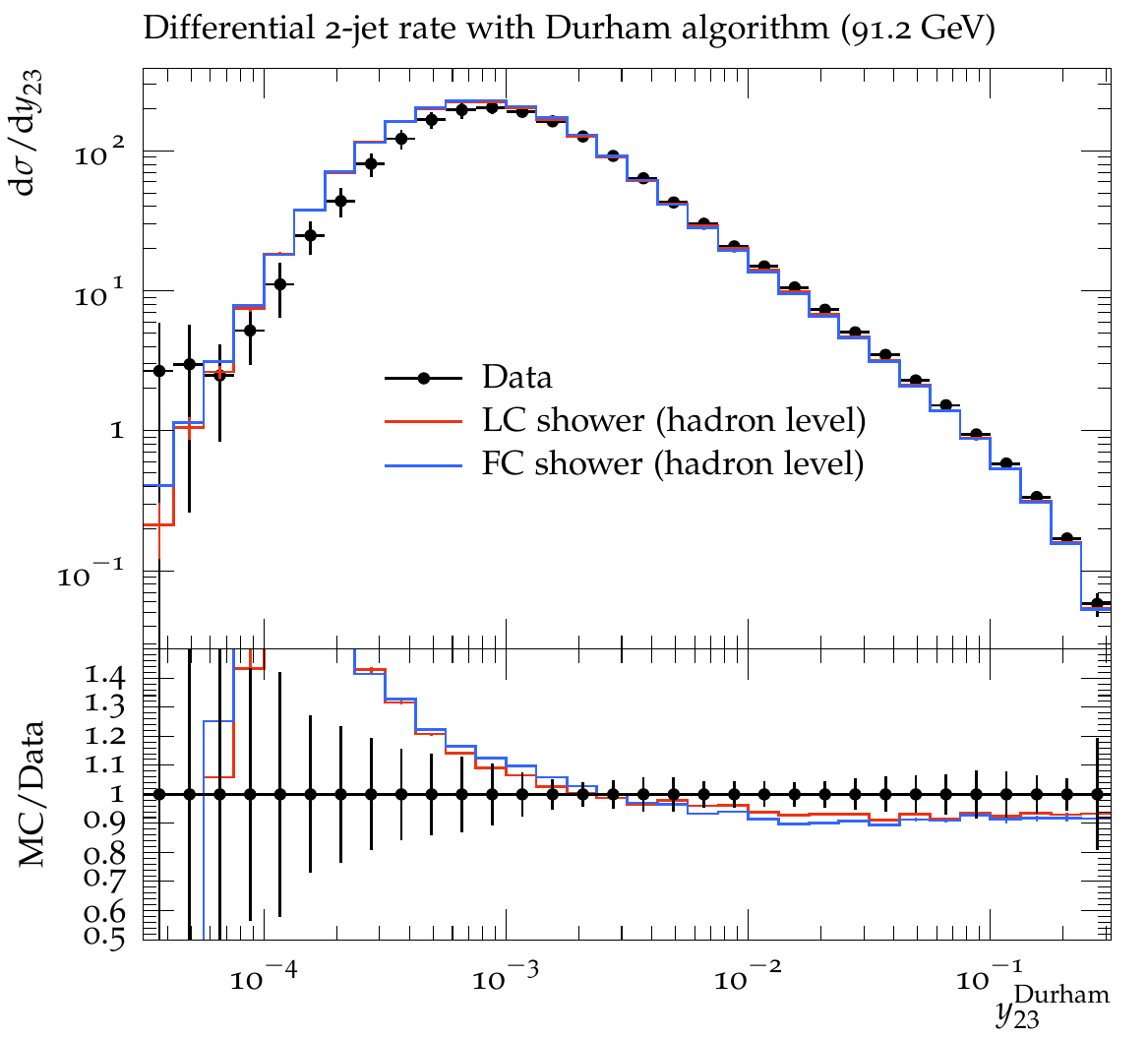}
    \caption{Jet separation between two- and three-jet configurations in the Durham algorithm, as measured by OPAL~\cite{Pfeifenschneider:1999rz} and 
    implemented in RIVET~\cite{Buckley:2010ar}, compared to leading color and fixed color parton showers.}
    \label{fig:Diff2JetHad}
    \end{minipage}
    \hfill
    \begin{minipage}[b]{0.45\textwidth}
    \includegraphics[width=\textwidth]{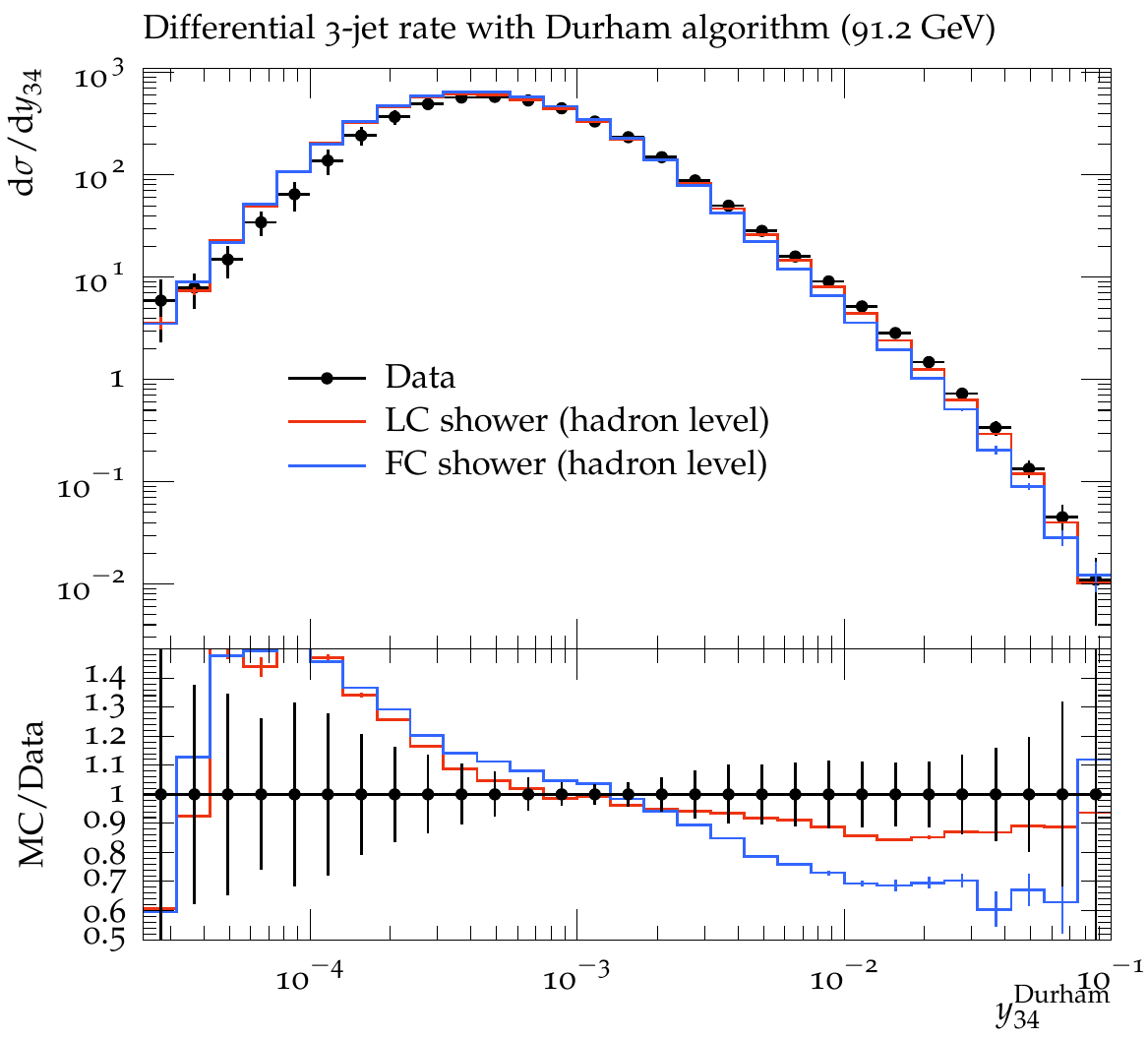}
    \caption{Jet separation between three- and four-jet configurations in the Durham algorithm, as measured by OPAL~\cite{Pfeifenschneider:1999rz} and 
    implemented in RIVET~\cite{Buckley:2010ar}, compared to leading color and fixed color parton showers.}
    \label{fig:Diff3JetHad}
    \end{minipage}
\end{figure}

Larger effects are anticipated for differential clustering scales. 
Figure~\ref{fig:Diff2JetHad} compares the jet separation between two- and 
three-jet events to OPAL measurements~\cite{Pfeifenschneider:1999rz}. The result of LC
and FC parton showers are very similar, since the observable is dominated by 
the first parton shower emission, which is identical in both algorithms.
A more drastic change can be found in the jet separation between three- and
four-jet states $y_{34}$ shown in Fig.~~\ref{fig:Diff3JetHad}. The impact
of the modified kinematics because of radiation from sub-leading color
dipoles produces a much softer spectrum, as already observed to a lesser degree
in Figs.~\ref{fig:Thrust} and~\ref{fig:JetBroadening}. Of course, it needs
to be noted that small differences between LC and FC showers in the 
non-perturbative region ($y_{34}\lesssim 5\cdot 10^{-4}$) can have significant
impact of $\mathcal{O}(10\%)$ also in the hard region because of the 
normalization of the plot. In general, we do however observe a
tendency towards softer multi-jet spectra in the FC parton shower. This 
is expected because of the inclusion of depleting sub-leading corrections. 
The kinematics of these corrections is, in our algorithm, completely determined 
by the splitting $|\mathcal{M}\rangle$, and not influenced by 
the potentially different propagator structure of splittings in 
 $\langle \mathcal{M}|$. The data comparison raises the question if subleading 
color corrections can really be treated independently from other, kinematic,
corrections. We believe that this study gives important input for future 
developments of more accurate parton showers.

\section{Conclusions and Outlook}\label{sec:conclusions}

\noindent
This articles has presented the first implementation of a fixed color
parton shower algorithm that remains numerically feasible and stable for
an arbitrary number of emissions. This has been achieved by leveraging
the qualities of the color flow basis. The growth in complexity has
been tamed by employing a stochastic sampling of color configurations. 
Remaining numerical instabilities (that could be ameliorated by accumulating
higher statistics) have been improved with the introduction of a cutoff
on fixed color evolution. This necessitates keeping track of auxiliary events
that allow a matching onto a leading color shower. We have also discussed
an infrared safe matching to the Lund string hadronization model. 

Preliminary comparisons to LEP data have been presented. Here, the data 
description should not be regarded as final -- in particular because
known higher-order corrections are not present to avoid double-counting
in the fixed color result -- but rather as allowing to assess
benefits of differences between leading- and fixed color evolution. These
comparisons indicate an unfavorable trend towards too soft radiation patterns
in the fixed color parton shower. This observation is an important input 
for efforts to define parton showers at higher accuracy, since it suggests that
the exponentiation of complete color correlators without also including
multi-parton kinematic correlations (beyond the three-particle
correlations given by Eq.~\ref{eq:CS}) might have undesirable consequences. 

The algorithm has been implemented as stand-alone \textsc{Python}  code that was 
interfaced to \textsc{Pythia} via Les Houches files. A natural next step is to extend
the fixed color evolution also to initial-state splittings. Due to the handling
of multiple parton interactions and beam remnants, this will require a more
native implementation within the event generator framework. Further logical
next steps would be the introduction of $\mathcal{O}(\alpha_s^2)$ corrections to
allow recovering the two-loop cusp-anomalous dimension, and the inclusion of 
multi-parton kinematic correlations. In the future it will be important to 
study the complex interplay of sub-leading color, kinematics, and higher order
corrections to better describe the data.

\begin{acknowledgments}
  \noindent
  We thank S.~H{\"o}che for help in developing the algorithm described in
  Sec.~\ref{sec:Algorithm} through numerous discussions, and for pointing out 
  the option to use an independent cutoff for fixed color evolution.
  We further thank J.~Forshaw and S.~Pl{\"a}tzer for interesting discussions
  on the Glauber gluons, and S.~Keppeler and M.~Sj{\"o}dahl for discussions on color bases.
  This work was supported by Fermi Research Alliance, LLC, under Contract 
  No.\ DE--AC02--07CH11359 with the U.S.\ Department of Energy, Office of 
  Science, Office of High Energy Physics.
\end{acknowledgments}

\appendix 

\begin{figure}
 \label{fig:AlgorithmFlowchart}
    \includegraphics[width=\textwidth]{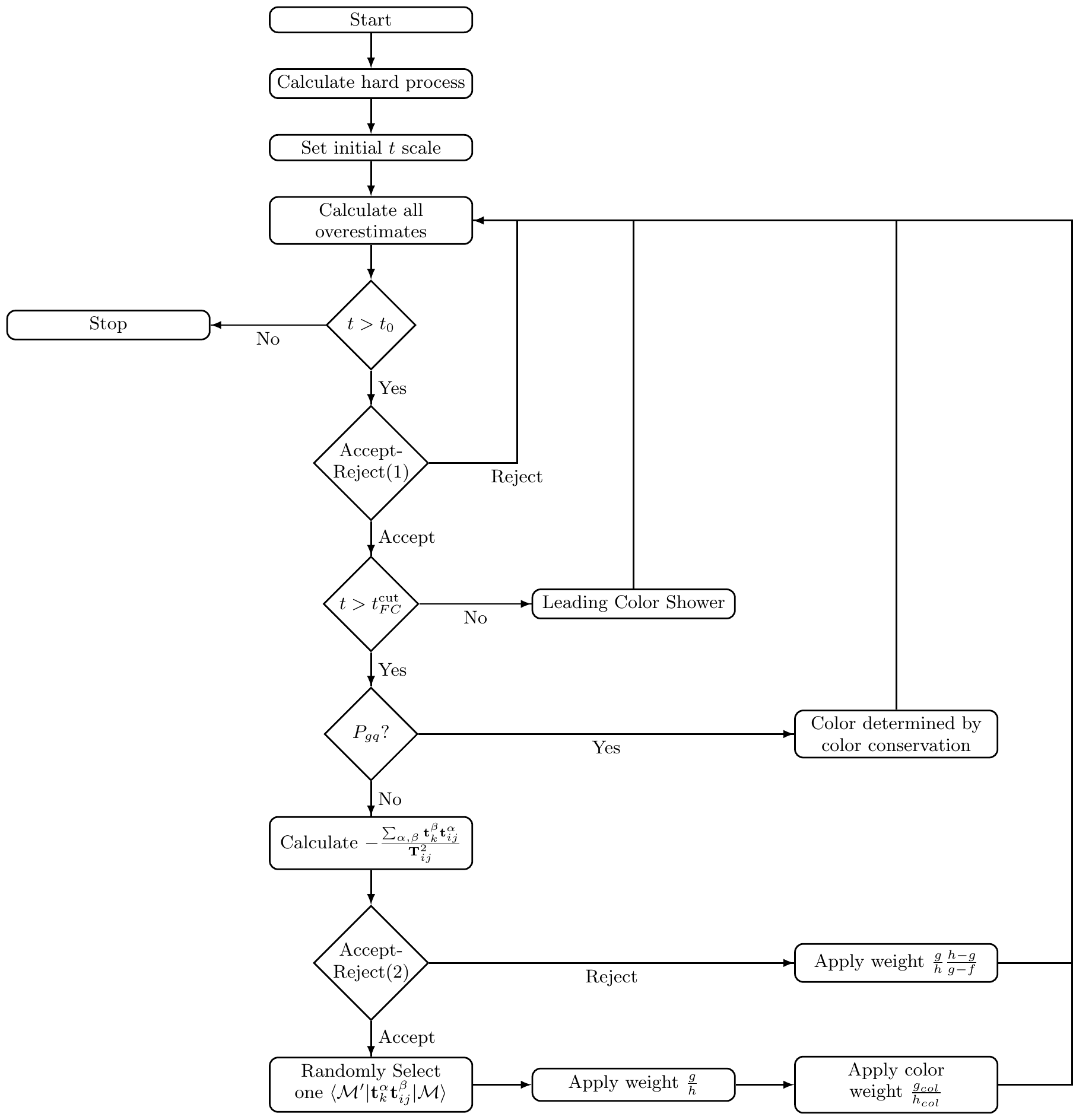}
    \caption{Flowchart representing the FC parton shower algorithm. Details on the calculation and of the definitions of $f$, $g$, $h$, $g_{col}$, and $h_{col}$ can be found in Sec.~\ref{subsec:alg} and Sec.~\ref{subsec:color_sampling}.}
\end{figure}

\section{The default tune of the \textsc{Dire} plugin for \textsc{Pythia}}\label{app:tune}

\noindent
Since the default tune of the \textsc{Dire} plugin to \textsc{Pythia} has not been documented 
beyond comments in the {\tt direforpythia} code repository and source code, we
here give the tune settings, as introduced in revision 
SHA\\ {\tt ca99627ea3c6741ddb98ea55054c44c6b298c0ba} of 
{\tt gitlab.com/dire/direforpythia}:
\begin{verbatim}
  StringPT:sigma             = 0.2952
  StringZ:aLund              = 0.9704
  StringZ:bLund              = 1.0809
  StringZ:aExtraDiquark      = 1.3490
  StringFlav:probStoUD       = 0.2046
  StringZ:rFactB             = 0.8321
  StringZ:aExtraSQuark       = 0.0
  TimeShower:pTmin           = 0.9
\end{verbatim}
These tune settings have been extracted by comparing against 
$e^+e^-\rightarrow$ hadrons data from $\sqrt{s}=14-91.2$ GeV, as well as
comparisons against Tevatron and LHC data. The following \textsc{Dire} plugin-specific
settings were assumed
\begin{verbatim}
  PDF:pSet                   = LHAPDF6:MMHT2014nlo68cl
  PDF:pHardSet               = LHAPDF6:MMHT2014nlo68cl
  TimeShower:alphaSvalue     = 0.1201
  SpaceShower:alphaSvalue    = 0.1201
  ShowerPDF:usePDFalphas     = on
  ShowerPDF:useSummedPDF     = on
  DireSpace:forceMassiveMap  = on
  ShowerPDF:usePDFmasses     = off
  DireTimes:kernelOrder      = 1
  DireSpace:kernelOrder      = 1
\end{verbatim}
This implies that the soft-collinear pieces of splitting kernels 
were rescaled with corrections to recover the inclusive $\mathcal{O}(\alpha_s^2)$
correction in the soft limit (see e.g.\ \cite{Bendavid:2018nar}).

\bibliographystyle{apsrev4-1}
\bibliography{ColorShower}

\end{document}